\documentclass[a4paper]{aa}
\usepackage[utf8]{inputenc}
\usepackage{amsmath}

\usepackage[]{natbib}
\bibpunct{(}{)}{;}{a}{}{,} 

\RequirePackage{xcolor}
\definecolor{lightgray} {gray}{0.7}
\definecolor{darkgray}  {gray}{0.43}
\definecolor{darkblue}  {rgb} {0,0,.65}
\definecolor{citegreen} {rgb} {0.28,0.41,0.19}
\definecolor{darkgreen} {rgb} {0.37,0.72,0.17}
\definecolor{lightgreen}{rgb} {0.56,0.88,0.28}
\RequirePackage[%
  pdfstartview=FitH,%
  breaklinks=true,%
  bookmarks=true,%
  colorlinks=true,%
  linkcolor= darkblue,
  anchorcolor=black,%
  citecolor=citegreen,
  filecolor=black,%
  menucolor=black,%
  urlcolor=darkblue%
  ]{hyperref}

\newcommand{\kms}{$\rm{km\,s}^{-1}$}

\newcommand{\Ho}{$H_{0}$}

\newcommand{\Hunit}{$\rm{km\,s^{-1}\,Mpc^{-1}}$}

\begin{document}

\titlerunning{Redshift-Leavitt Bias and the Hubble Constant}
\title{Leavitt Laws Applied to Dilated Variability Periods Bias the Distance Scale}
\title{Towards a 1\% measurement of the Hubble Constant: accounting for time dilation in variable-star light curves}

\author{Richard I. Anderson\thanks{ESO fellow, \email{randerso@eso.org}}}
 \institute{European Southern Observatory, Karl-Schwarzschild-Str. 2, D-85748 Garching b. M\"unchen, Germany}
\date{Received 27 August 2019 / Accepted 23 September 2019}

\abstract{Assessing the significance and implications of the recently established Hubble tension requires the comprehensive identification, quantification, and mitigation of uncertainties and/or biases affecting $H_0$ measurements. Here, we investigate the previously overlooked distance scale bias resulting from the interplay between redshift and Leavitt laws in an expanding Universe:  Redshift-Leavitt bias (RLB). Redshift dilates oscillation periods of pulsating stars residing in supernova-host galaxies relative to periods of identical stars residing in nearby (anchor) galaxies. Multiplying dilated $\log{P}$ with Leavitt Law slopes leads to underestimated absolute magnitudes, overestimated distance moduli, and a systematic error on $H_0$. Emulating the {\it SH0ES} distance ladder, we estimate an associated $H_0$ bias of $(0.27 \pm 0.01)\,\%$ and obtain a corrected $H_0 = 73.70 \pm 1.40$\,\Hunit.
RLB becomes increasingly relevant as distance ladder calibrations pursue greater numbers of ever more distant galaxies hosting both Cepheids (or Miras) and type-Ia supernovae. The measured periods of oscillating stars can readily be corrected for heliocentric redshift (e.g. of their host galaxies) in order to ensure $H_0$ measurements free of RLB.}


\maketitle

\section{Introduction}
Pulsating stars such as classical Cepheids enable precise measurements of the local expansion rate of the Universe, $H_0$, thanks to the existence of period-luminosity relations (PLRs), or Leavitt laws \citep[henceforth: LLs]{Leavitt1908,Leavitt1912}. Using a Cepheids-based distance ladder, the {\it SH0ES} team \citep[henceforth: R+16]{Riess2016} recently established a systematic difference between the present-day value of $H_0$ and the value inferred based on Cosmic Microwave Background observations by the \citet{Planck2018H0} assuming the concordance cosmological model (flat $\Lambda$CDM). Following further improvements to the distance ladder \citep[e.g.][]{Riess2018,Riess2019}, and considering Cepheid-independent routes to measure $H_0$ \citep[e.g.][]{Wong2019,Freedman2019,Yuan2019}, the so-called Hubble tension now figures at a significance of $4-6\sigma$, sparking an increasing number of suggested modifications to $\Lambda$CDM \citep[for a detailed overview, see][and references therein]{Verde2019}.

The {\it SH0ES} distance ladder consists of three rungs that are fitted globally (cf. Appendix in R+16). The first rung involves the calibration of the Cepheid LL in relatively nearby, so-called ``anchor'' galaxies whose distances are known. The second rung consists of so-called ``SN-host'' galaxies, where both Cepheids and type-Ia supernovae (SNe\,Ia) have been observed: SN-host galaxies thus set the luminosity zero-point for SNe\,Ia. The third, final rung consists of the Hubble diagram of SNe\,Ia. The intercept of this Hubble diagram provides the measurement of $H_0$. Almost all SN-host galaxies are more distant than the anchor galaxies, with the exception of M\,101, which is approximately $0.9$\,Mpc closer than the anchor galaxy NGC\,4258. Following an improvement of the distance to NGC\,4258, the latest $H_0 = 73.5 \pm 1.4$\Hunit\ \citep{Reid2019}. 

Following impressive gains in $H_0$ precision and the importance of $H_0$ for informing modifications to $\Lambda$CDM, the identification and mitigation of previously overlooked $H_0$ uncertainties and biases is rapidly gaining  importance. Notably, the effects of stellar association bias \citep{AndersonRiess2018} are now taken into account in the $H_0$ measurement \citep{Riess2019}.

A key bottleneck for increasing $H_0$ precision is the number of SN-host galaxies. At present,  Cepheids can be measured with good precision in galaxies up to $\sim 40$\,Mpc distant (R+16), and large efforts are in progress to increase the number of SN-host galaxies from currently 19 to 38 \citep{Riess2019}. However, the rather low volumetric rate of SNe\,Ia explosions requires exploring alternative primary distance indicators capable of probing greater distances, such as Mira-variable stars \citep[e.g.][]{Whitelock2008,Yuan2017,Huang2018,Huang2019,Bhardwaj2019}, especially considering the short mission duration of the James Webb Space Telescope ({\it JWST}). Adding new SN-host galaxies to the distance ladder will therefore tend to include more distant galaxies and increase average SN-host galaxy redshift due to the Hubble-Lema\^itre law \citep{Wirtz1924,Lemaitre1927,Hubble1929}.

The longitudinal Doppler effect shifts the oscillation frequency emitted by a source as a function of its radial (line-of-sight) velocity, $v_r$. Time dilation due to cosmological redshift $\bar{z}$ slows down clocks in a mathematically approximately identical way in the case of small $\bar{z}$ and non-relativistic velocities ($v_r \ll c$). Thus, time dilation of the type Ia supernova SN\,1995K enabled observational proof of the Universe's expansion \citep{Leibundgut1996}.
Several studies have considered the impact of inaccuracies related to redshift measurements of SNe\,Ia \citep[e.g.][]{Hui2006,Davis2011,Davis2019} on $H_0$ and the dark energy equation of state. Additionally, the impact of Doppler frequency shifts for asteroseismic inferences was studied by \citet{Davies2014} and is gaining importance for measuring orbital motion in pulsating stars \citep{Murphy2014}. 
However, the impact of systematic differences in (observed heliocentric) redshift between anchor and SN-host galaxies on variable-star periods and the $H_0$ measurement has hitherto remained unexplored. 

This \emph{article} investigates the relevance and impact of dilated variable-star periods on the measurement of $H_0$. Specifically, \S\ref{sec:RLB} explains the distance bias arising from cosmic expansion, the Doppler effect, and the Leavitt law in the case of Cepheids and Mira stars. \S\ref{sec:H0} estimates the impact on the latest $H_0$ measurement involving Cepheids and SNe\,Ia. \S\ref{sec:discussion} discusses these findings in the context of future distance ladders based on Mira-variable stars in the era of the {\it JWST}. Finally, \S\ref{sec:conclusions} summarises this work and presents its conclusions.

\section{Redshift-Leavitt Bias (RLB)}\label{sec:RLB}
\begin{figure}
    \centering
    \includegraphics{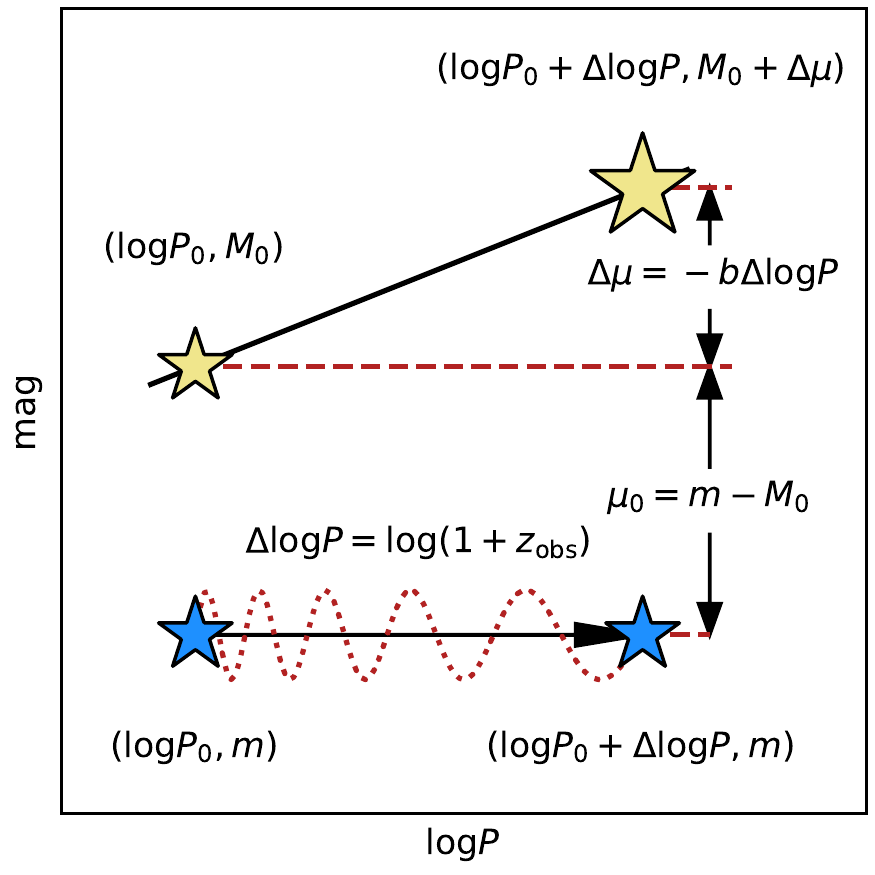}
    \caption{Not-to-scale visual representation of Redshift-Leavitt Bias. SN-hosts are redshifted relative to anchor galaxies. Observed oscillation frequencies of Cepheids, Miras, and other variable stars, are subject to time dilation, i.e., observed periods are longer than emitted periods. This translates to a distance modulus bias due to the slope of Leavitt laws.\label{fig:cartoon}}
\end{figure}

The observed total redshift of an extragalactic source relative to the observer is a combination of cosmological redshift ($\bar{z}$), peculiar motion of the galaxy ($z_{\rm{pec}}^{\rm{gal}}$), peculiar motion of the observer ($z_{\rm{pec}}$), and gravitational redshift of the galaxy ($z_{\rm{\phi}}^{\rm{gal}}$) and the observer ($z_{\rm{\phi}}$). Once corrected to the heliocentric reference frame, the observed redshift $z_{\rm{obs}}$ is \citep[their Eq.\,2.2]{Calcino2017}:
\begin{equation}
1 + z_{\rm{obs}} = (1 + \bar{z})(1 + z_{\rm{pec}}^{\rm{gal}})(1 + z_{\rm{pec}})(1 + z_{\rm{\phi}}^{\rm{gal}})(1+ z_{\rm{\phi}}) \ .
\label{eq:redshift}
\end{equation}

Variability periods of extragalactic variable stars such as classical Cepheids are subject to both time dilation due to cosmological redshift (mostly relevant for Cepheids outside the Local Group), and the Doppler effect due to line-of-sight motions such as peculiar velocities, velocity dispersions, partial sampling of a galaxy's rotation curve, orbital motion, etc. Since the latter velocities are usually $\lesssim 1000$\,\kms, such Doppler shifts can be treated non-relativistically to very good approximation. The observed redshift of an individual variable star in another galaxy is thus:
\begin{equation}
1 + z_{i,\rm{obs}} \approx (1 + z_{\rm{obs}})\left(1 + \frac{v_{r,i}}{c}\right)\ ,
\label{eq:redshiftdoppler}
\end{equation}
where $v_{r,i}$ is the line-of-sight component of the star's motion relative to the heliocentric reference frame, not counting peculiar motion, which is already included as part of Eq.\,\ref{eq:redshift}. Of course, $v_{r,i}$ of individual Cepheids is currently measurable only in the Milky Way and Magellanic Clouds, where cosmological redshift is negligible. For individual variable stars in other galaxies, internal motions due to velocity dispersion and partially sampled rotation curves can lead to $v_{r,i}/c \lesssim 10^{-3}$. However, most observed populations of Cepheids in SN-host galaxies sample all parts of their host galaxy disks, which furthermore tend to be oriented relatively face-on. Hence, no significant net effect is to be expected for a typical SN-host galaxy's Cepheid population as a whole, meaning, $\langle v_{r} \rangle \approx 0$. For partially sampled galaxies \citep[e.g. the PHAT footprint of M31, cf.][]{Dalcanton2012}, disk or halo rotation could lead to a net $\langle v_r \rangle \lesssim \pm 300$\,\kms, which would be comparable to the effect of $\bar{z}$ at distances $\lesssim 4$\,Mpc.

Galaxy catalogs list redshifts ($z_{\rm{obs}}$) measured as the displacement of spectral lines relative to their rest-wavelength $\lambda_0$, corrected to the heliocentric reference frame \citep[e.g.][]{Huchra1992}:
\begin{equation}
z_{\rm{obs}} = \frac{\lambda - \lambda_0}{\lambda_0} \ .
\label{eq:redshiftmeasured}
\end{equation}
$z_{\rm{obs}}$ does not distinguish between the physical origin of redshift, and includes all terms listed in Eq.\,\ref{eq:redshift}. However, it does not account for motions of individual stars ($v_{r,i}$) or velocity differences across a galaxy. In the following, we neglect these latter two contributions since Cepheid populations are usually distributed across the full disks of SN-host galaxies, which are usually oriented face-on (R+16). However, observers may consider accounting for net effects, due to galaxy rotation at distances $\lesssim 40$\,Mpc, if these conditions are not fulfilled.

\begin{figure}
    \centering
    \includegraphics{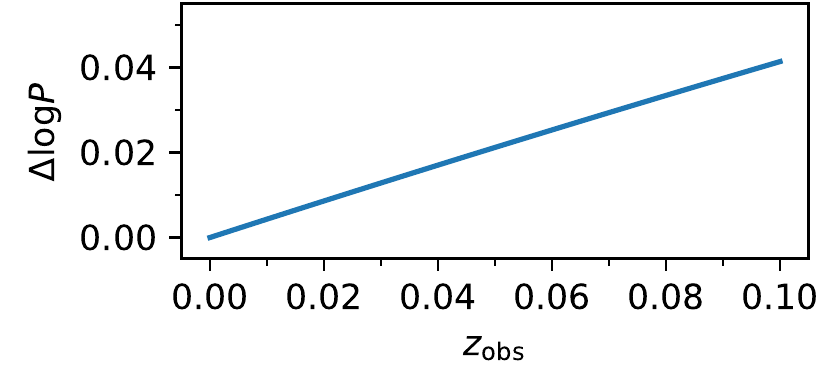}
    \caption{\label{fig:timedilation} Dilation of logarithmic oscillation period as a function of redshift, cf. Eq.\,\ref{eq:deltalogP}.}
    \vspace{0.2cm}
    \includegraphics{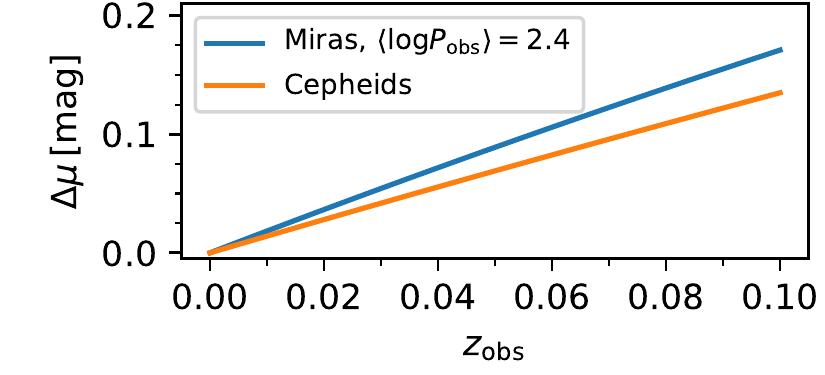}
    \caption{\label{fig:distmod} Redshift-Leavitt Bias for Cepheids and Miras with $\langle \log{P_{\rm{obs}}} \rangle = 2.4$ whose LL is calibrated at $\log{P_{\rm{ref}}}=2.3$, cf. Eqs.\,\ref{eq:leavitt}, and \ref{eq:leavittMira}.}
\end{figure}

Using $z_{\rm{obs}}$, we calculated the effect of time dilation on variable-star periods (cf. Fig.\,\ref{fig:timedilation}) as
\begin{equation}
P_{\rm{obs}} = (1 + z_{\rm{obs}})P_{0} \ ,\ \mathrm{so\ that}
\label{eq:TimeDilation}
\end{equation}
\vspace{-0.5cm}
\begin{equation}
\Delta \log{P} = \log{P_{\rm{obs}}} - \log{P_0} = \log{(1 + z_{\rm{obs}})} = \log{B}\ ,
\label{eq:deltalogP}
\end{equation}
where we use $\log{B} = \log{(1 + z_{\rm{obs}})}$ for brevity.

Leavitt Laws relate oscillation periods of certain pulsating star types to  absolute magnitudes $M$. The most common functional form is linear in $\log{P}$ and is used, for example, for classical Cepheid variables:
\begin{equation}
M = a + b\cdot\log{P} \ ,
\label{eq:leavitt}
\end{equation}
where $a$ and $b$ are calibrated using objects at known distances (ideally from geometry), for example, via {\it Gaia} parallaxes \citep{GaiaDR2Brown,Lindegren2018}, detached eclipsing binaries in the LMC \citep{Pietrzynski2019}, or the parallax of the mega-maser in NGC\,4258 \citep{Humphreys2013}. 

Non-linear LLs have also been considered, for example, broken PLRs for Cepheids \citep[e.g.][]{Bhardwaj2016}, PLRs affected by metallicity \citep[e.g.][especially for RR Lyrae stars observed in the infrared]{Sesar2017,Gieren2018,Delgado2019}. LLs are noticeably non-linear for Mira stars, where quadratic LLs  have been used \citep[e.g.][]{Yuan2017,Huang2018}:
\begin{equation}
M = a + b_1 \cdot (\log{P_{\mathrm{obs}}} - 2.3) + b_2 \cdot (\log{P_{\mathrm{obs}}} - 2.3)^2 \ .
\label{eq:leavittMira}
\end{equation}

In the following, we adopt $b=-3.26$ for a linear LL appropriate for Cepheids ($H-$band Wesenheit PL-slope from R+16) and $b_1 = -3.59$, $b_2 = -3.40$, and $\log{P_{\rm{ref}}} = 2.3$ \citep{Huang2018} for quadratic LLs. 

Absolute magnitudes inferred using LLs, $M_{\rm{LL}}$, are biased by the effect of time dilation on oscillation periods, cf. Fig.\,\ref{fig:cartoon}. With Eq.\,\ref{eq:deltalogP} and Eqs.\,\ref{eq:leavitt} \& \ref{eq:leavittMira}, we obtain
\begin{equation}
\begin{split}
\Delta M & = M_{\rm{LL}} - M_0 \\ 
 & = b\Delta\log{P} = b\log{B} & \small{[\rm{lin\, LL}]}\\ 
 & = b_1  \log{B} + b_2  \log{B}  \left( 2 \log{P_{\rm{obs}}^{'}}  - \log{B} \right) & \small{[\rm{quad\, LL}]} & \ ,
\end{split}
\label{eq:MLL}
\end{equation}
where the second line applies to linear LLs and the third line to quadratic LLs. For linear LLs, the bias depends only on $b$ and the observed heliocentric redshift. For quadratic LLs, RLB depends on $z_{\rm{obs}}$, $b_1$, and $b_2$, as well as the pivot period $P_{\rm{ref}}$, since $\log{P_{\rm{obs}}^{'}} = \log{P_{\rm{obs}}} - \log{P_{\rm{ref}}}$.

The resulting bias in distance modulus is therefore:
\begin{equation}
\Delta \mu = \mu_{\rm{LL}} - \mu_0 = m - M_0 - \Delta M - m + M_0 = -\Delta M \ .
\label{eq:deltamu}
\end{equation}
Inserting Eq.\,\ref{eq:MLL} in Eq.\,\ref{eq:deltamu} we obtain:
\begin{equation}
\begin{split}
\Delta \mu = & - b\log{B} & \small{\rm{[lin\, LL]}} \\
\Delta \mu = & - b_1  \log{B} - b_2  \log{B}  \left( 2 \log{P_{\rm{obs}}^{'}}  - \log{B} \right) & \small{\rm{[quad\, LL]}} &\ .
\end{split}
\label{eq:deltamuLL}
\end{equation}
Figure\,\ref{fig:distmod} illustrates $\Delta\mu$ for a wide range of $z_{\rm{obs}}$. We notice that a) $(\log{B})^2 \sim 10^{-4}$ is negligible at distances $< 100$Mpc, and b) $\log{P_{\rm{obs}}^{'}}$ can be minimized by using a pivot $\log{P_{\rm{ref}}}$ close to the sample average. However, any galaxy is expected to show a distribution of Mira periods, so that time dilation could lead to a small apparent LL slope change in a given galaxy if redshift is not accounted for.

With $\mu = 5\log{d} + 25$ ($d$ in Mpc), the ratio of the true distance $d_0$ to the biased distance $d_{\rm{LL}}$ is:
\begin{equation}
\frac{d_0}{d_{\rm{LL}}} = 10^{-0.2\cdot\Delta \mu} = 10^{0.2\cdot\Delta M} \ .
\label{eq:drelbias}
\end{equation}
With Eq.\,\ref{eq:deltamuLL}, Eq.\,\ref{eq:drelbias} becomes:
\begin{equation}
\begin{split}
d_0 &= d_{\rm{LL,lin}} \cdot 10^{0.2 b\log{B} } \\
d_0 &= d_{\rm{LL,quad}} \cdot 10^{0.2 \left[  b_1 \log{B} + b_2  \log{B}  \left( 2 \log{P_{\rm{obs}}^{'}}  - \log{B} \right) \right] } 
\end{split}
\label{eq:drelbiasLL}
\end{equation}
for linear and quadratic LLs, respectively. Since $b$, $b_1$, and $b_2$ are all negative, and $\log{B} \ge 1$, this typically means that $d_0 < d_{\rm{LL}}$. We notice that the term $\log{P_{\rm{obs}}^{'}}$ can be negative or positive, depending on pivot period and average $\log{P_{\rm{obs}}}$. In the following, we consider the small difference between $\log{P_{\rm{ref}}}=2.3$ and $\langle \log{P_{\rm{obs}}} \rangle = 2.4$, which corresponds to the case of Miras in NGC\,4258 \citep[`Gold' sample in their Tab.\,6]{Huang2018}.

\begin{figure}
    \centering
    \includegraphics{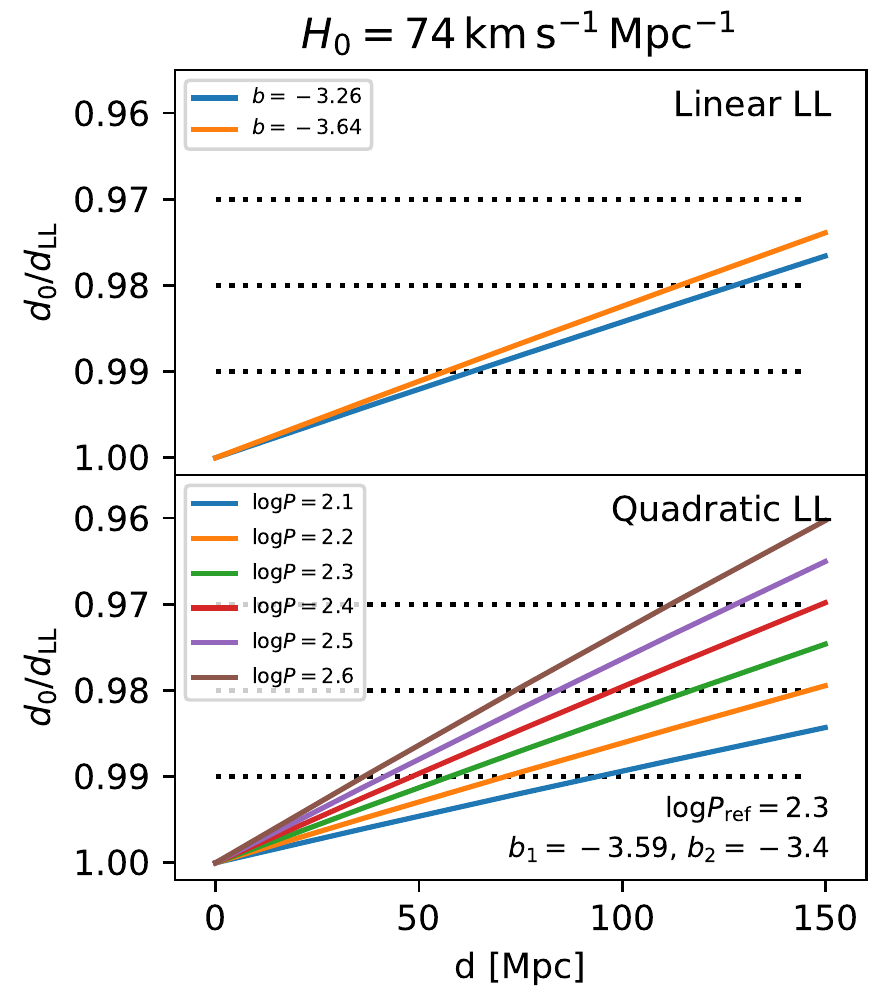}
    \caption{\label{fig:reldisterr} Relative distance error as function of distance in case of  cosmological redshift (considers only expansion). Relative distance errors of $1$, $2$, and $3\%$ are shown by horizontal dotted lines. For Cepheids, we adopt $b = -3.26$ (corresponds to the $H$-band Period-Wesenheit relation using F160W, F555W, and F814W in R+16), for Miras $b_1 = -3.59$, $b_2 = -3.40$, and $\log{P_{\rm{ref}}}=2.3$ from \citet{Huang2018}. If left uncorrected, RLB affects Cepheids at the $1\%$ ($2\%$) level for $d > 62$\,Mpc ($d > 123$\,Mpc). For Miras of mean $\log{P_{\rm{obs}}}=2.4$, the $1\%$ ($2\%$) bias level is reached sooner, at $d > 47.5$\,Mpc ($d > 95$\,Mpc).}
\end{figure}

Figure\,\ref{fig:reldisterr} illustrates RLB as a function of distance assuming $H_0 = 74$\,\Hunit\ and redshift caused by cosmic expansion only. The slightly steeper slopes of Mira LLs imply a slightly stronger susceptibility to RLB compared to Cepheids. For quadratic LLs, more significant differences  result from possible differences among the average observed and pivot period (if $\log{P_{\rm{obs}}^{'}} \ne 0$).

$H_0$ is measured as the intercept of the Hubble diagram, $a_x$ (cf. Eqs.\,5 and 9 in R+16, subscript $x$ denotes photometric band) via
\begin{equation}
\log{H_0} = \frac{M_x^0 + 5a_x + 25}{5} \ ,\ \mathrm{which\ yields}
\label{eq:H0meas}
\end{equation}
With Eq.\,\ref{eq:H0meas} and $\Delta M = M_{x,\rm{obs}}^0 - M_{x,\rm{true}}^0$, we obtain:
\begin{equation}
\begin{split}
H_{0,\rm{true}} &= H_{0,\rm{LL,lin}} \cdot 10^{-0.2 b\log{B} } \\
H_{0,\rm{true}} &= H_{0,\rm{LL,quad}} \cdot 10^{-0.2 \left[  b_1 \log{B} + b_2  \log{B}  \left( 2 \log{P_{\rm{obs}}^{'}}  - \log{B} \right) \right] } 
\end{split}
\label{eq:corrH0}
\end{equation}
for linear and quadratic LLs, respectively, where $H_{0,\rm{true}} > H_{0,\rm{LL}}$ since $b$, $b_1$, $b_2$ all $< 0,$ and $\log{B} = \log{(1 + z_{\rm{obs}})} > 0$. Hence, the Universe is expanding slightly faster than previously reported based on the Cepheid-based distance ladder \citep[R+16,][]{Riess2018,Riess2019,Reid2019}. In \S\ref{sec:H0}, we estimate this effect and the correction for $H_0$, and underline the importance of correcting variability periods for time dilation when measuring distances exceeding $40$\,Mpc using Mira stars with {\it JWST}.

\section{Correcting $H_0$ for RLB}\label{sec:H0}

We now estimate RLB as it applies to the {\it SH0ES} Cepheids-based distance ladder (R+16) and the corresponding $H_0$ measurement. To this end, we compiled measured heliocentric redshifts from NED\footnote{\url{https://ned.ipac.caltech.edu}} as well as other relevant information for the anchor and SN-host galaxies. Table~\ref{tab:table} provides this information for convenience. 

\begin{table*}
\centering
\caption{\label{tab:table} Information used to estimate redshift-Leavitt bias for anchor and SN-host galaxies in SH0ES Cepheids-based distance ladder (R+16). The number of Cepheids in each galaxy, $N_{\rm{Cep}}$, approximate distance modulus $\mu$, and PL-relation dispersion $\sigma_{\rm{PL}}$ are all taken from R+16. Heliocentric  redshifts $z_{\rm{obs}}$ are compiled from NED, with original references $\rm{Ref}_{z}$ as follows:
a: \citet{Vaucouleurs1991},
b: \citet{Meyer2004,Wong2006},
c: \citet{Koribalski2004},
d: \citet{Bureau1996},
e: \citet{Krumm1980},
f: \citet{Guthrie1996},
g: \citet{Verheijen2001},
h: \citet{Lauberts1989},
i: \citet{Kent2008},
j: \citet{Grogin1998},
k: \citet{Strauss1992},
l: \citet{Schneider1992},
m: \citet{Richter1987}.
The redshift for the anchor galaxies and SN-host galaxies, weighted by $\sigma_{\rm{PL}}$, are shown above each group. $\Delta z_{\rm{obs,RF}} = z_{\rm{obs}} - z_{\rm{obs,cal}}$ is the redshift difference between each SN-host galaxy and the anchor reference frame. The shift in logarithmic oscillation period, $\Delta \log{P}$ (cf. Eq.\,\ref{eq:deltalogP}), the overestimate of distance modulus $\Delta \mu$ (cf. Eq.\,\ref{eq:deltamuLL}), and the distance bias $\Delta d$ (cf. Eq.\,\ref{eq:drelbias}) are computed using $\Delta z_{\rm{obs,RF}}$.
}
\begin{tabular}{lrrrrrrrrr}
\hline
Galaxy & $N_{\rm{Cep}}$  & $\mu$ & $\langle \sigma_{\rm{PL}} \rangle$ & $z_{\rm{obs}}$ & Ref$_{z}$ & $\Delta z_{\rm{obs,RF}}$ & $\Delta \log{P}$ & $\Delta \mu$ & $\Delta d = d_{0} - d_{\rm{LL}}$ \\
 & & [mag] & [mag] & [$10^{-3}$] & & [$10^{-3}$] & [d] & [mmag] & [Mpc] \vspace{3mm}\\
\hline
\multicolumn{10}{c}{Anchor galaxies, $\langle z_{\rm{obs,cal}} \rangle = 0.48\times 10^{-3}$} \\
\hline
Milky Way & 50  &  $-$    & 0.08 & $0$    &   & $-$ & $-$ & $-$ & $-$ \\
LMC       & 785 &  18.477 & 0.12 & $0.92$ & m & $-$ & $-$ & $-$ & $-$ \\
NGC\,4258 & 139 &  29.397 & 0.15 & $1.49$ & a & $-$ & $-$ & $-$ & $-$ \\
\hline
\multicolumn{10}{c}{SN-host galaxies, $\langle z_{\rm{obs,LL}} \rangle = 4.76\times 10^{-3}$} \\
\hline
M\,101     & 251  &  29.135  & 0.32  & $0.80$ &  a & $0.32$  & 0.007 & 0.2  & -0.001 \\
NGC\,1015  & 14   &  32.497  & 0.36  & $8.77$ &  b & $8.29$  & 0.353 & 11.5 & -0.168 \\
NGC\,1309  & 44   &  32.523  & 0.36  & $7.12$ &  c & $6.64$  & 0.281 & 9.2  & -0.135 \\
NGC\,1365  & 32   &  31.307  & 0.32  & $5.45$ &  d & $4.97$  & 0.209 & 6.8  & -0.057 \\
NGC\,1448  & 54   &  31.311  & 0.36  & $3.90$ &  c & $3.41$  & 0.141 & 4.6  & -0.039 \\
NGC\,2442  & 141  &  31.511  & 0.38  & $4.89$ &  b & $4.41$  & 0.184 & 6.0  & -0.056 \\
NGC\,3021  & 18   &  32.498  & 0.51  & $5.14$ &  a & $4.66$  & 0.195 & 6.4  & -0.093 \\
NGC\,3370  & 63   &  32.072  & 0.33  & $4.27$ &  e & $3.78$  & 0.157 & 5.1  & -0.061 \\
NGC\,3447  & 80   &  31.908  & 0.34  & $3.56$ &  f & $3.07$  & 0.127 & 4.1  & -0.046 \\
NGC\,3972  & 42   &  31.587  & 0.38  & $2.84$ &  g & $2.36$  & 0.095 & 3.1  & -0.030 \\
NGC\,3982  & 16   &  31.737  & 0.32  & $3.70$ &  a & $3.21$  & 0.133 & 4.3  & -0.044 \\
NGC\,4038  & 13   &  31.290  & 0.33  & $5.48$ &  h & $4.99$  & 0.210 & 6.8  & -0.057 \\
NGC\,4424  & 3    &  31.080  & 0.56  & $1.46$ &  i & $0.97$  & 0.035 & 1.2  & -0.009 \\
NGC\,4536  & 33   &  30.906  & 0.29  & $6.03$ &  j & $5.55$  & 0.234 & 7.6  & -0.053 \\
NGC\,4639  & 25   &  31.532  & 0.45  & $3.40$ &  b & $2.91$  & 0.120 & 3.9  & -0.036 \\
NGC\,5584  & 83   &  31.786  & 0.33  & $5.46$ &  c & $4.98$  & 0.209 & 6.8  & -0.072 \\
NGC\,5917  & 13   &  32.263  & 0.38  & $6.35$ &  k & $5.87$  & 0.248 & 8.1  & -0.106 \\
NGC\,7250  & 22   &  31.499  & 0.43  & $3.89$ &  l & $3.41$  & 0.141 & 4.6  & -0.042 \\
UGC\,9391  & 28   &  32.919  & 0.43  & $6.38$ &  l & $5.90$  & 0.249 & 8.1  & -0.144 \\
\hline
\end{tabular}
\end{table*}

We first determine the difference between the mean redshift of LL anchor galaxies and the mean redshift of SN-host galaxies used for measuring $H_0$. The $\sigma_{\rm{PL}}$-weighted average observed redshift of anchor galaxies is $\langle z_{\rm{obs,cal}} \rangle \approx 0.48 \times 10^{-3}$. For Milky Way Cepheids, we adopt $v_r = 0$\,\kms,\ as expected for a random distribution of radial velocities. Any deviations from null velocity are on the order of $10$\,\kms\ and can be comfortably neglected. 
For SN-host galaxies, the $\sigma_{\rm{PL}}$-weighted average redshift is $\langle z_{\rm{obs,LL}} \rangle \approx 4.76 \times 10^{-3},$ and the average redshift difference between anchor and SN-host galaxies is $\Delta z_{\rm{obs}} = 4.28 \times 10^{-3}$. In the following, we use $\log{B} = \log{(1 + \Delta z_{\rm{obs}})}$ to calculate RLB and the bias of $H_0$.

\begin{figure}
    \centering
    \includegraphics{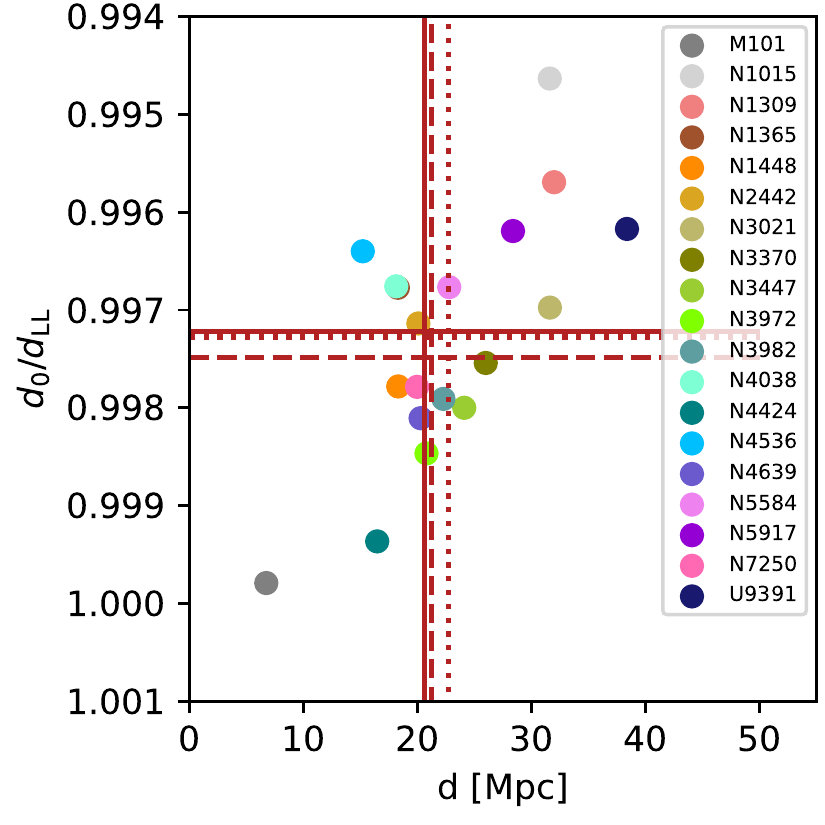}
    \caption{\label{fig:measuredRLB}Relative distance error incurred due to RLB, cf. Eq.\,\ref{eq:drelbias}. $z_{\rm{obs}}$ was translated to distance assuming $H_0=74$\,\Hunit; scatter in the relation arises from peculiar velocities. The dark red solid, dashed, and dotted crosses indicate the average values obtained for different weightings. Solid lines: weighted according to $\sigma_{\rm{PL}}$ in Tab.\,\ref{tab:table}, $\langle d_0 / d_{\rm{LL}} \rangle = 0.9972$. Dashed lines: weighted according to $N_{\rm{cep}}$ in Tab.\,\ref{tab:table}. Dotted lines: unweighted mean distance and bias, $\langle d_0 / d_{\rm{LL}} \rangle = 0.9975$.}
\end{figure}

Figure\,\ref{fig:measuredRLB} shows  $d_0 / d_{\rm{LL}}$ for all 19 {\it SH0ES} SN-host galaxies individually, as well as sample averages including the $\sigma_{\rm{PL}}$-weighted average (dark red solid lines), the average weighted by number of Cepheids, $N_{\rm{cep}}$ (dashed), and the un-weighted average (dotted). For the $\sigma_{\rm{PL}}$-weighted average, we find $\langle d_0/d_{\rm{LL}} \rangle_{\sigma_{\rm{PL}}} \approx 0.99721$. Using the $N_{\rm{Cep}}$-weighted average, we find $\langle d_0/d_{\rm{LL}} \rangle_{N_{\rm{Cep}}} \approx 0.99748$. Any uncertainty contribution due to redshift uncertainties is minimal, since $\langle \sigma_{z}/z \rangle = 0.004$ restricts variations in $\Delta \log{P}$ to $\lesssim 10^{-5}$.

For $H_0 = 73.50 \pm 1.40$\,\Hunit\ \citep{Reid2019} and $\Delta z_{\rm{obs}}=4.28\times10^{-3}$, Eq.\,\ref{eq:corrH0} yields $H_{0,\rm{true}} = 73.70 \pm 1.40$\,\Hunit, i.e., an increase of $\Delta H_{0} = 0.20 \pm 0.01$\,\Hunit. Thus, RLB amounts to $14\%$ of the reported total uncertainty on $H_0$ of $1.90\%$. The small shift in $H_0$ slightly increases the significance of the tension between {\it Planck} and the {\it SH0ES} distance ladder from $4.1\sigma$ to $4.2\sigma$. 

Analogously, we estimate biases of $0.19\%$ and $0.20\%$ for recent Mira-based \Ho\ measurements by \citet{Huang2019} that used linear Mira LL slopes of $b =-3.64$ and $-3.35$. In this case, NGC4258 was the sole anchor, and NGC1559 \citep[$z_{\rm{obs,N1559}}=4.35 \times 10^{-3}$]{Koribalski2004} the sole SN-host galaxy. 

Of course, the above is a somewhat crude, first-order estimation of the degree by which previous $H_0$ measurements are affected by RLB. Future $H_0$ measurements should take into account the dilation of observed oscillation periods using Eq.\,\ref{eq:deltalogP} to avoid RLB. \S\ref{sec:discussion} highlights why this correction is required to elucidate Hubble tension using a future distance ladder based on more distant Mira stars.

\section{Discussion}\label{sec:discussion}
Time dilation can affect any oscillating star's variability period and lead to distance bias whenever there is a net redshift difference among the LL calibration set and the population where LLs are being applied. Besides the aforementioned Cepheid, RR Lyr, and Mira stars, many other classes of pulsating stars, including $\delta$~Scuti stars, 
type-II Cepheids, 
and long-period oscillating red giants obey PLRs that render them potentially useful as standard candles \citep[e.g.][]{Ziaali2019,Matsunaga2011,Kiss2003}. 

Mira stars are of particular importance due to their high luminosity and large amplitudes for future distance ladder calibration using {\it JWST}. Mira stars observed with {\it JWST} should significantly extend the distance $d$ within which SNe\,Ia luminosity be cross-calibrated. Since volume increases as $d^3,$ and the probability of a SN\,Ia exploding depends on volume, it follows that most new SN-host galaxies observed with {\it JWST} should reside at significantly greater distance than the current {\it SH0ES} SN-host galaxies. Due to the Hubble-Lema\^itre law, the redshift difference among anchor and SN-host galaxies should therefore be enhanced in the era of the Mira-{\it JWST} distance ladder.

Fig.\,\ref{fig:reldisterr} illustrates this effect, and shows that RLB of a sample of Miras with $\langle \log{P_{\mathrm{obs}}} \rangle = 2.5\ (316\,\rm{d})$  exceeds $2\%$ at Coma cluster distances of $100$\,Mpc. Moreover, the intra-cluster velocity dispersion $\sigma_{v_{\rm{pec}},\rm{Coma}} \approx 1000$\,\kms\ \citep[and references therein]{Sohn2017} could differentially bias distances of Coma cluster galaxies by $1-2\%$, 
potentially affecting the interpretation of galaxy cluster scales. 

According to Eq.\,\ref{eq:drelbiasLL}, a mismatch between the average period of a sample of Miras and the pivot period of quadratic LLs can further increase RLB.
Assuming $\log{P_{\rm{ref}}}=2.3$, a bias of $1\%$ is reached at $37$, $42$, $48$, $57$, $71$, and $93$\,Mpc for locally calibrated Miras of $\log{P} = 2.6$, $2.5$, $2.4$, $2.3$, $2.2$, and $2.1$, respectively. Alternatively expressed, at $60$\,Mpc, individual Mira stars would be biased by between $0.7\%$ and $1.7\%$, depending on their period. 

RLB depends on LL slopes, which depend on photometric passbands. For example, the $H-$band LL in R+16 has slightly shallower slope ($b_H = -3.06$) than the near-IR Wesenheit LL ($b_W = -3.26$). Hence, the $H-$band Cepheid LL is less affected by RLB than the near-IR Wesenheit-LL. Analogously, different kinds of oscillating stars exhibiting shallower LL slopes are also less strongly affected by RLB. For example, the slope of the period term in the $K-$band RR Lyrae PL-metallicity relation has slope of approximately $-2.4$ \citep[e.g.][]{Barth2002,Minniti2003}, reducing the effect compared to Cepheids or Miras.

Thankfully, RLB is easily avoided by correcting observed oscillation periods for time dilation effects. In the continued pursuit of measuring $H_0$ with $1\%$ accuracy, this effect can and must be accounted for.

\section{Conclusions}\label{sec:conclusions}
The Universe's expansion leads to a subtle, distance-dependent dilation of variability periods in oscillating stars. If left uncorrected, this systematic change in variability period biases distance estimates based on Leavitt laws along the distance ladder, because more distant stars are assumed to be more luminous than they truly are. Due to the interplay between redshift and Leavitt law slopes, we term this effect Redshift-Leavitt Bias (RLB). 

RLB results in overestimated distances and thus leads to an underestimated value of the Universe's local expansion rate $H_0$. Emulating the {\it SH0ES} distance ladder \citep{Riess2016,Riess2018,Riess2019,Reid2019}, we estimate a bias of $\Delta H_0 / H_0 = 0.27 \pm 0.01\%$. Applying this first-order correction to the  $H_0$ reported by \citet{Reid2019}, we obtain $\Delta H_0 = 0.20 \pm 0.01$\,\Hunit\ and a de-biased $H_0 = 73.70 \pm 1.40$\,\Hunit. The slight increase of $H_0$ increases the significance of the Hubble tension between the ``early-universe'' value by the \citet[$H_0 = 67.4 \pm 0.5$\,\Hunit]{Planck2018H0} and the Cepheids-based distance ladder to $4.2\sigma$.

With oscillating stars being observed at increasing distances, correcting variable-star oscillation periods for time dilation due to redshift becomes increasingly important. For the highly promising Mira stars, we estimate $H_0$ bias of order $2-3\%$ if this effect is not accounted for. Hence, a future Mira-based distance ladder requires correcting variability periods for time dilation.

\begin{acknowledgements} 
The author is pleased to thank the anonymous referee for a timely and constructive report, Adam G. Riess and Stefano Casertano for useful comments on an earlier draft, as well as Bruno Leibundgut, Jason Spyromilio, and Steven Kawaler for useful discussions.
This research has made use of NASA's Astrophysics Data System.
\end{acknowledgements}

\bibliographystyle{aa} 
\bibliography{biblio}

\end{document}